\documentclass[aps,prb,twocolumn,showpacs,a4paper,unsortedaddress,
amsmath,amssymb,byrevtex]{revtex4}

\usepackage[latin1]{inputenc}
\usepackage{graphicx}
\usepackage{dcolumn}
\usepackage{bm}
\usepackage{times}

\begin{document}

\preprint{ffuov/05-04}

\title{Strongly correlated electron physics in nanotube-encapsulated
metallocene chains}

\author{V. M. Garc\'{\i}a-Su\'arez$^1$}
\author{J. Ferrer$^2$}
\author{C. J. Lambert$^1$}

\affiliation{$^1$ Department of Physics, Lancaster University,
Lancaster, LA1 4YB, U. K.}

\affiliation{$^2$ Departamento de F\'{\i}sica, Universidad de
Oviedo, 33007 Oviedo, Spain}


\date{\today}

\begin{abstract}

The structural, electronic and transport properties of metallocene
molecules (MCp$_2$) and isolated  or nanotube-encapsulated
metallocene chains are studied by using a combination of density
functional theory and non-equilibrium Green's functions. The
analysis first discusses the whole series of isolated MCp$_2$
molecules, where M = V, Cr, Mn, Fe, Co, Ni, Ru, and Os. The series
presents a rich range of electronic and magnetic behaviors  due to
the interplay between the crystal field interaction and Hund's
rules, as the occupation of the $d$ shell increases. The article
then shows how many of these interesting properties can also be
seen when MCp$_2$ molecules are linked together to form periodic
chains. Interestingly, a large portion of these  chains display
metallic, and eventually magnetic, behavior. These properties may
render these systems as useful tools for spintronics applications
but this is hindered by the lack of mechanical stability of the
chains. It is finally argued that encapsulation of the chains
inside carbon nanotubes, that is exothermic for radii larger than
4.5 \AA, provides the missing mechanical stability and electrical
isolation. The structural stability, charge transfer, magnetic and
electronic behavior of the ensuing chains, as well as the
modification of the electrostatic potential in the nanotube wall
produced by the metallocenes are thoroughly discussed. We argue
that the full devices can be characterized by two doped, strongly
correlated Hubbard models whose mutual hybridization is almost
negligible. The charge transferred from the chains produces a
strong modification of the electrostatic potential in the nanotube
walls, which is amplified in case of semiconducting and
endothermic nanotubes. The transport properties of isolated
metallocenes between semi-infinite nanotubes are also analyzed and
shown to lead to important changes in the transmission
coefficients of clean nanotubes for high energies.
\end{abstract}


\maketitle

\section{Introduction}

Carbon nanotubes \cite{Iji91} have been for long proposed as
possible elements for electronic circuits on the nanoscale. Apart
from their mechanical stability and small diameter, these devices
can be produced in large quantities and be easily assembled.
Furthermore, they can be shaped in very different geometric
configurations which make them specially suitable to join
different parts of a nanocircuit and achieve novel switching
features\cite{Ban05}. On the other hand the electronic properties
can range from insulating to metallic \cite{Dre04} depending on
the radius and the chirality. In the absence of defects, electron
transport falls in the ballistic regime and therefore large
carrier mobility, high yield and low power consumption is
expected. Additional functionality can be achieved by promoting
defects in the nanotube walls\cite{Post01}, or by doping them.
{\em n}- and {\em p}-doped nanotubes can be fabricated by pulling
charge in or out of the walls
\cite{Kaz99,Zho00,Col00,Jou00,Kong01,Mar01,Kaz01,Shi01,Jav02,Der02}.
These functionalities enable the nanotubes to mimic different
parts of a transistor\cite{Tan98,Yao99,Post01} and even to produce
devices with negative differential resistance \cite{Zho00}.

Several methods have been used to produce {\em n}- and {\em
p}-doped SWNTs but most of them have the problem that the
resulting SWNTs are not stable upon air exposition due to
oxidation processes which tend to {\em p}-dope the carbon wall. A
better procedure consists of encapsulating \cite{Tsa94} metallic
atoms \cite{Gue94}, or even organic molecules like fullerenes
\cite{Hor02} inside the nanotube walls. Metallic elements and
compounds tend to form continuous conducting nanowires
\cite{Tsa94}, while fullerenes produce local distortions in the
electronic band structure of the nanotube \cite{Hor02}. This way,
fullerenes can be used to section the nanotube into different
segments \cite{Lee02}. This local change of the gap should
effectively produce a series of quantum dots that could localize
electrons\cite{Cho03}. Trapping single electrons may be a used to
produce entangled electron pairs and also to tailor interactions
between static and flying qubits \cite{Gun06}. The type of charge
transfer can be traced back to electron affinity and the
ionization potential of the encapsulated molecules. It is
recognized now that molecules with high electron affinity tend to
remove charge from the nanotube wall and produce {\em p}-doped
SWNTs, whereas molecules with low ionization potential tend to
transfer charge to the wall and produce {\em n}-doped
SWNTs\cite{Tak03,Lu04}.

Nanotubes can also be rendered magnetically active upon
encapsulation of magnetic elements or organometallic molecules.
These new devices offer the possibility to separate the magnetic
material from the environment and hence reduce its oxidation.
Additionally, encapsulation may give rise to new magnetic
properties not present in conventional magnetic nanowires
\cite{Pic04} and even change the mechanical stability of the
nanotubes \cite{Kar04}. These include, for example, hysteresis
loop shifts due to exchange bias \cite{Pra02,Kar05}, changes in
the complex permittivity and permeability \cite{Che04}, increased
coercitivity and Barkhausen jumps \cite{Sat02}, magnetic
anisotropy \cite{Muh02}, size and number-dependent interactions
between catalyst particles \cite{Jan04} and many other magnetic
phenomena. Other fascinating properties include the disappearance
of magnetism in small clusters of Ni atoms \cite{Yag04} or the
antiferromagnetic behavior shown by iron \cite{Wei04,Wei06}.

The subject-matter of this article is a theoretical proposal for
the encapsulation of specific metallorganic molecules in
nanotubes, that combine the two functionalities sketched in the
former two paragraphs, e. g.: tailored doping and addition of
magnetic behavior to carbon nanotubes. The proposed metallorganic
molecules are called metallocenes \cite{Tog98,Lon98} and are
composed of a metallic transition metal (TM) atom and two
cyclopentadyenil (Cp) rings made out of five carbon and five
hydrogen atoms each. The configuration of both rings is specially
stable if they loose an electron, thus transforming in aromatic
rings where the $\pi$-electrons are delocalized around the ring.
This produces a negative compound that can bind to another Cp ring
through a TM atom with a valence of +2 or higher. The case with a
valence of +4, which is specially relevant for the fabrication of
polymers is not dealt with in the present paper.  The article
instead focuses only on the simplest metallocene molecule
structures, which have only two Cp rings bridged by the TM atom.
We note that some of these molecules have already been suggested
as the functional core of ingenious electronics devices. For
instance, FeCp$_2$ (ferrocene) has recently been proposed as a
molecular bridge that can rival the conductance of conjugated
molecules \cite{Eng01,Get05} and as a molecular switch
\cite{Liu06}. A spintronics switch made out of two CoCp$_2$
molecules (cobaltocenes), sandwiched by two magnetic leads, has
also been proposed \cite{Liu05}. Explicit encapsulation of
cobaltocene in carbon nanotubes was recently realized by Li and
coworkers \cite{Li05}, who showed that such composite devices are
stable for nanotube diameters of 9.4 \AA. For thicker nanotubes
only a modified version of the cobaltocene
(bis(ethylcyclopentadienyl)) was observed to remain inside the
tube. This was possibly due to the fact that cobaltocenes, which
are smaller than bis(ethylcyclopentadienyl), bonded too weakly to
the wall of the thicker tubes and were expelled from them at the
slightest thermal or mechanical fluctuations. Li and coworkers
also reported a charge transfer from the molecule to the nanotube
wall and a red shift in the photoluminescence spectra due to the
formation of localized impurity states, which could make these
systems specially suitable to produce {\em n}-doped nanotubes and
modulate the electronic properties of the nanotube wall.

Noting that TMCp$_2$ molecules could be synthesized with TM=V, Cr,
Mn, Co and Ni, as well as Ru and Os, and that each of them display
different electronic and magnetic properties, we proposed
\cite{Gar06} to use them as off-the-shell elements to fabricate
one-dimensional chains of metallocene molecules, encapsulated
inside the carbon nanotubes walls. In other words, we proposed to
fabricate nanoscale wires whose tailored electronic, magnetic and
conducting properties would be due to the transition metal atoms
of the metallocene molecules, and where the major role of the
nanotube would be to actually provide mechanical and electric
isolation to the metallocene core. We envisioned two possible
categories of devices. In the first class, the nanotube walls
would also be metallic, and therefore the properties of the device
would come from a non-conventional interplay between two
non-trivial electron systems. In the second category, the nanotube
walls remain semiconducting, and the analog of a electrical
appliance wire on the nanoscale would be realized.

In this article we take further our previous calculations and show
a host of additional features that can be present in these
systems. We analyze the structural and magnetic properties of
isolated metallocene molecules and isolated or encapsulated
chains. We confirm that the devices display a rich variety of
electronic properties that can indeed be traced back to the
diverse filling of the metallocene molecular levels. We find that
the metallocene chains are very accurate realizations of the
one-dimensional Hubbard model, with hopping, Hubbard $U$ and
doping parameters that can be extracted unequivocally from the
simulations. These chains transfer charge to the nanotube walls,
but do not hybridize with them. For the metallic nanotube class we
therefore find two Hubbard liquids that act as reservoirs of each
other. On the contrary, for the semiconducting nanotube class, and
whenever the encapsulation is endothermic (e.g.: when the diameter
of the nanotube is small), we find that the metallocene molecules
produce significant energy barriers on the nanotube walls that
could localize electrons. We believe that these devices may be
useful for quantum computation, due to the possibility to realize
and tailor interactions of localized and flying qubits
\cite{Gun06}.

The outline of the article is as follows: we describe in Section
\ref{SecII} our assembly of theoretical methods. The properties of
isolated metallocene molecules and chains of metallocenes are
shown in Section \ref{SecIII}. In Sections \ref{SecIV} and
\ref{SecV} we focus on the properties of chains of metallocenes
inside nanotubes and the mapping to strongly correlated models. In
Section \ref{SecVI} we study the transport properties of isolated
metallocenes inside carbon nanotubes. We close the article with a
section that presents and discusses our conclusions.

\section{Theoretical Method \label{SecII}}

The calculations were based on Density Functional Theory
\cite{Koh65} as implemented in the first-principles code SIESTA
\cite{Sol02}, which uses norm-conserving pseudopotentials
\cite{Tro91} to get rid of the core electrons and linear
combinations of localized atomic orbitals $\phi_\mu(\vec r-\vec
d_\mu)$ to span the valence states \cite{San89}. The
pseudopotentials, which were optimized so that the local part was
smooth \cite{Sol02}, also included non-linear core corrections
\cite{Lou82} for the transition metal atoms of the MCp$_2$
molecules, with core correction radii of 0.70 a. u. These were
obtained by following the recipe used to generate the
pseudopotential in iron explained in Ref. \onlinecite{Gar04a}. The
basis set had a double$-\zeta$ character for carbon and hydrogen
atoms and double$-\zeta$ polarized for the transition metal atoms.
The cutoff radii were determined by minimizing the energy for a
given reference system, i. e.: the H$_2$ molecule for hydrogen, a
diamond crystal for carbon, and the bulk phase for the
corresponding transition metal atom.

Due to the localized character of the atomic orbitals special care
had to be taken to minimize the impact on our results of both the
basis set superposition error (BSSE) \cite{Boy70} and the basis
set incompletion error (BSIE) \cite{Gle96,Her98}. The first error
comes from the fact that the Hilbert space of the composite system
is different from the Hilbert space of the each part taken
separately and therefore the binding energy is smaller due to an
additional contribution which comes from the the presence of more
localized states. In order to correct BSSE, it is necessary then
to subtract the additional contribution by calculating in separate
simulations the energy of each individual system in the geometry
of the composite system, but inserting ghost states at the
location of the missing atoms. Then, the binding energy of the
device is estimated as

\begin{equation}
\Delta E=\Delta E_0-(E_{A+GB}-E_A)-(E_{B+GA}-E_B)
\end{equation}

\noindent where $A$ and $B$ represent the nanotube and the
molecule in the relaxed configuration of the composite system,
$GA$ and $GB$ are the corresponding ghost states and $\Delta
E_0=E_{AB}-E_{A_\mathrm{rel}}-E_{B_\mathrm{rel}}$ is the binding
energy calculated without ghost states and relaxing all
structures.

The contribution from the BSIE appears when there is large
interstitial separation between different parts of a system which
is described with localized basis sets. Although it is not
possible to estimate this contribution it is known to have an
opposite sign to the BSSE. This implies that the calculated values
of the BSSE-corrected binding energy are upper bounds. We note
anyway that we were careful enough to span all the interstitial
space between the molecule and the nanotube wall so that the
states of the molecule and the nanotube wall had enough overlap.

The exchange and correlation energy and potential were calculated
with the generalized gradient approximation (GGA) as parameterized
by Perdew, Burke and Ernzerhof \cite{Per96}, since this
approximation works much better for transition metal elements
\cite{Gar04a,Gar04b}. The real space grid used to compute the
integrals leading to density and Hamiltonian matrix elements was
calculated by setting a plane wave cutoff of 200 Ry. Finally, we
took a number of $k$-points along the direction of the nanotube
($z$-axis), ranging from 20 to 100, depending on the size of the
system and on wether the nanotube was semiconducting or metallic.

The transport properties were obtained with the ab-initio code
SMEAGOL \cite{Roc06}, which uses the Hamiltonian provided by
SIESTA to calculate the retarded Green's function of the
scattering region (M), which in the basis of atomic orbitals is
expressed as

\begin{equation}
G^\mathrm{R}_{\mathrm{MM},\mu\nu}(E)=[E^+S-H-
\Sigma^\mathrm{R}_\mathrm{L}(E)-
\Sigma^\mathrm{R}_\mathrm{R}(E)]^{-1}_{\mathrm{MM},\mu\nu}
\end{equation}

\noindent where $S$ is the overlap matrix, $H$ is the Hamiltonian,
$E^+=E+i\delta$ and

\begin{eqnarray}
\Sigma^\mathrm{R}_\mathrm{L}(E)=[E^+S-H]_\mathrm{ML}
G^\mathrm{R}_\mathrm{LL}(E)[E^+S-H]_\mathrm{LM}\\
\Sigma^\mathrm{R}_\mathrm{R}(E)=[E^+S-H]_\mathrm{MR}
G^\mathrm{R}_\mathrm{RR}(E)[E^+S-H]_\mathrm{RM}
\end{eqnarray}

\noindent are the self-energies that take into account the
coupling to the left and right semi-infinite electrodes.
$G_\mathrm{LL(RR)}$ are the surface Green's functions. From the
retarded Green's function and its Hermitian conjugate, the
advanced Green's function $G^\mathrm{A}$, it is possible to obtain
the density matrix using the Keldysh formalism \cite{Kel65}, so
that

\begin{equation}
\rho_{\mu\nu}=\frac{1}{2\pi
i}\int_C\mathrm{d}EG^<_{\mathrm{MM},\mu\nu}(E)
\end{equation}

\noindent where $G^<$ is the retarded Green's function, which in
general is calculated as

\begin{equation}
G^<_\mathrm{MM}(E)=G(E)^\mathrm{R}_\mathrm{MM}
[\Gamma_\mathrm{L}f_\mathrm{L}+
\Gamma_\mathrm{R}f_\mathrm{R}](E)G(E)^\mathrm{A}_\mathrm{MM}
\end{equation}

\noindent where
$\Gamma_\mathrm{L(R)}=i[\Sigma^\mathrm{R}_\mathrm{L(R)}-
\Sigma^\mathrm{A}_\mathrm{L(R)}]$ are the gamma matrices that take
into account the strength of the coupling to the leads and
$f_\mathrm{L(R)}(E)=f(E-\mu_\mathrm{L(R)})$ are the Fermi
distribution functions of the leads. Such a procedure is necessary
to obtain the electronic density, $n(\vec
r)=\sum_{\mu,\nu}\rho_{\mu\nu}\phi_\mu(\vec r-\vec
d_\mu)\phi_\nu(\vec r-\vec d_\nu)$, because the calculation deals
with semi-infinite leads and also in a general non-equilibrium
situation the distribution function of the scattering region is
not well defined and therefore the common diagonalization
procedure can not be used.

Once the self-consistency is achieved the conductances for spin-up
and spin-down electrons are calculated by using the familiar
expression \cite{Fis81}

\begin{equation}
{\cal G}_\sigma(E)=\frac{e^2}{h}\mathrm{tr}
[\Gamma_\mathrm{L}G^\mathrm{A}_\mathrm{MM}\Gamma_\mathrm{R}
G^\mathrm{R}_\mathrm{MM}\Gamma_\mathrm{R}]_\sigma(E)
\end{equation}

\noindent where $\sigma=\{\uparrow,\downarrow\}$ is the spin
index. The current is simply the integral of the conductance
multiplied by the difference of the Fermi distribution functions
of the leads.

\section{Isolated TMCp$_2$ molecules and chains \label{SecIII}}

As stated above, metallocenes where the transition metal belong to
the 3$d$ row or the iron column (ruthenocene and osmocene) are
composed of a metallic atom sandwiched by two Cp rings. Such rings
can be located just one on top of the other, so that the axis that
passes through the middle of the rings and the metallic atom is
$C_5$ and therefore the symmetry is an eclipsed $D_{5h}$, or
rotated 36 degrees relative to each other, so that the symmetry is
$D_{5d}$, i. e. staggered, and each carbon or hydrogen atom sits
on top of one of the edges of the other ring. In many cases the
same metalloecene can have both structures depending on the phase.
For instance, vanadocene has $D_{5d}$ symmetry in the solid state,
but $D_{5h}$ in the gas phase. However, we always found in our
calculations the first configuration to be more stable than the
second one by an amount that could be as large as 0.03 eV (for the
cobaltocene). This difference can be due to the BSIE, that
artificially decreases the interaction between the rings. We have
anyhow tested for several of the calculations that both geometric
arrangements of the Cp rings produce almost identical results. We
are therefore confident that this difference is not important for
our study.

The theoretical average distance between carbon atoms in the
Cp$_2$ rings is 1.46$\pm 0.05$ \AA, although it can change
slightly depending on the type of metallocene. These results
compare rather well with experiments, which give an average
distance of 1.44 \AA$\,$ for the ferrocene\cite{Lon98}, for
example. The distance between carbon and hydrogen atoms is equal
to 1.10 \AA. The main structural parameter, which is the distance
between TM and carbon atoms is listed in table \ref{Tab01}, along
with the distance between the TM atom and the plane of the Cp
ring. In almost all cases the first type of distance is the same
for each of the ten carbon atoms in each metallocene. However, the
axis of the two Cp rings in chromocene and, to a lesser extent, in
cobaltocene, were found to be tilted, in agreement with
experiments, possibly due to the lifting of degeneracy by a
Jahn-Teller distortion \cite{Eic89}. The distances in these cases
vary between 2.09 and 2.26 \AA$\,$ for CrCp$_2$, and between 2.07
and 2.15 \AA$\,$ for CoCp$_2$. Table \ref{Tab01} shows in these
two case the average value. As can be seen the agreement between
theory and experiment is excellent, with the only exception of
manganocene. The reason of such discrepancy could be due to the
fact that the distances for MnCp$_2$ correspond to the high spin
state while we only considered the low spin configuration.

\begin{figure}
\includegraphics[width=\columnwidth]{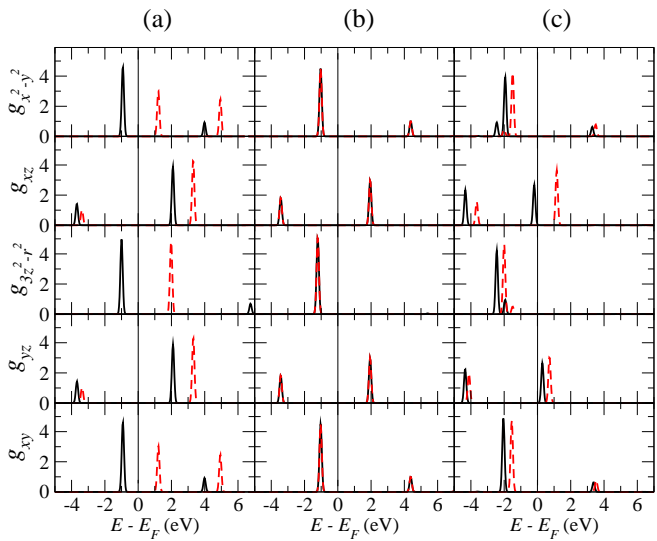}
\caption{\label{Fig1} Densities of States $g_i$ of (a) VCp$_2$,
(b) FeCp$_2$ and (c) CoCp$_2$, projected onto the $d$-shell
orbitals of the 3$d$-row metal atom, $i=d_{x^2-y^2}, d_{xz},
d_{3z^2-r^2}, d_{yz}, d_{xy}$. Continuous and dashed lines
represent spin-up and spin-down electrons, respectively.}
\end{figure}

The differences in the electronic structure of the metallocenes
can be traced back to the progressive filling of the $d$-shell of
the TM atom, that we also show in table \ref{Tab01}. These
electronic structures are determined by the interplay among
electron filling, Hund's rules and the crystal field interaction.
To illustrate this, we show in Fig. (\ref{Fig1}) the density of
states projected onto the five $d$-orbitals of the TM atom, of
VCp$_2$, FeCp$_2$ and CoCp$_2$. We find, first, that the $d_{xz}$
and $d_{yz}$ orbitals feel a different crystal field than the rest
of orbitals, since they are oriented towards the carbon atoms in
the Cp rings. They also overlap strongly with the carbon orbitals
and, consequently, their energy levels are split into bonding and
antibonding states. These two orbitals are degenerate in energy
for all metallocenes, see Fig. (\ref{Fig1} (a), (b)). Chromocene
and cobaltocene are an exception since, due to the Jahn-Teller
distortion the axis of the molecule is bent. Consequently, the
degeneracy of the $d_{xz}$ and $d_{yz}$ orbitals is lifted, as
shown in Fig. (\ref{Fig1} (c)). The energy levels of the
$d_{x^2-y^2}$, $d_{xy}$ and $d_{3r^2-z^2}$ are almost degenerate
and lie lower in energy because they are not oriented towards the
carbon atoms. Additionally, $d_{x^2-y^2}$ and $d_{xy}$ orbitals
overlap a little with carbon orbitals, while the $d_{3r^2-z^2}$
does not. Iron $d$-orbitals in FeCp$_2$ thereby arrange in a
closed shell structure; consequently FeCp$_2$ has no unpaired
electrons and is non-magnetic. The remaining metallocenes display
a sort of mirror symmetry about the closed-shell ferrocene. Notice
that the exchange splittings, shown in Fig. (\ref{Fig1}), are
proportional to the number of unpaired electrons. It is also
interesting to note that the geometric arrangement is such that
the orbital angular moment is quenched, and therefore the measured
values of the magnetization almost coincide with the spin moments.

\begin{table}
\caption{Theoretical and experimental distances between the TM and
the carbon atoms, between the carbon atoms and the Cp$_2$ rings,
and number of unpaired electrons. Experimental values from Ref.
\onlinecite{Lon98}.} \label{Tab01}
\begin{ruledtabular}
\begin{tabular}{lccccc}
\multicolumn{1}{l}{Metallocene}&\multicolumn{2}{c}{Distance TM-C
(\AA)}&\multicolumn{2}{c}{Distance TM-Cp
(\AA)}&\multicolumn{1}{c}{Unpaired
$e^-$}\\
&This study&Exp.&This study&Exp.&\\
\hline
VCp$_2$&2.27&2.28&1.91&1.92&3\\
CrCp$_2$&2.15 (Av.)&2.17&1.79 (Av.)&1.67&2\\
MnCp$_2$&2.07&2.38&1.66&2.14&1\\
FeCp$_2$&2.04&2.06&1.63&1.66&0\\
CoCp$_2$&2.11 (Av.)&2.12&1.71 (Av.)& &1\\
NiCp$_2$&2.20&2.20&1.82& &2\\
RuCp$_2$&2.21& &1.83&&0\\
OsCp$_2$&2.21& &1.83&1.86&0\\
\end{tabular}
\end{ruledtabular}
\end{table}

\begin{figure}
\includegraphics[width=\columnwidth]{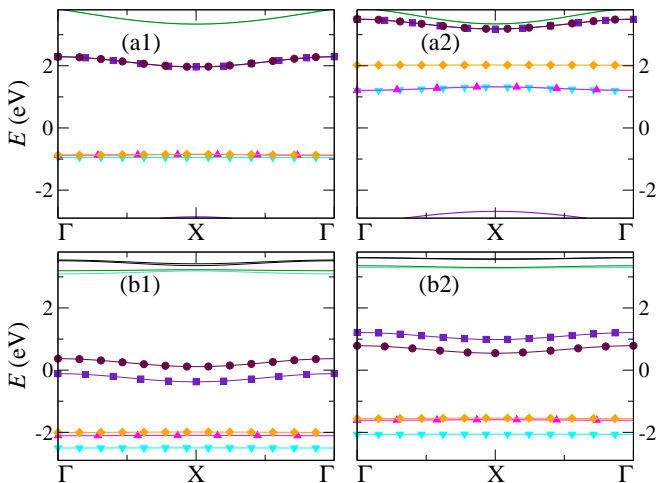}
\caption{\label{Fig2} Band structure around the Fermi level for
chains of vanadocene ((a1) and (a2) for majority and minority
spin, respectively) and cobaltocene ((b1) and (b2)) calculated at
a lattice constant of 7.5 \AA. Circles, squares, diamonds,
triangles up and triangles down represent $d_{yz}$, $d_{xz}$
($e_{1g}^*$ orbitals), $d_{x^2-y^2}$, $d_{xy}$ ($e_{2g}$ orbitals)
and $d_{3z^2-r^2}$ ($a_{1g}'$ orbital), respectively.}
\end{figure}

The distribution of electrons in the $d$ orbitals can also explain
why the distance between the Cp ring and TM atoms has a minimum
for ferrocene. We note that the same behavior is found for the
lattice constant of periodic isolated metallocene chains
\cite{Gar06}, where the $C_5$ axis of the molecule and of the
chain lie parallel to each other. The electronic band structure of
isolated VCp$_2$ and CoCp$_2$ chains is plotted in Fig.
(\ref{Fig2}). Notice that VCp$_2$ chains are semiconducting with a
large band gap. Moreover, their bands are essentially flat. This
is expected, since metallocene molecules are fairly stable and
should show small overlaps among each other. The band structure of
CoCp$_2$ chains is more interesting. First, the band width is
larger -of about a quarter of an electron volt- and, second
CoCp$_2$ chains are semiconducting, as expected, but the gap is
fairly small. Both chains are strong ferromagnets, since their
exchange splittings are such that up- and down-spin bands do not
overlap in energy. CoCp$_2$ chains are therefore more promising
for fabrication of transport devices, and this paper shall mostly
focus on them henceforth.


We remind that CoCp$_2$ molecules possess a bent axis, that leads
to a lifting of degeneracy of the molecular levels associated to
the $d_{xz}$ and $d_{yz}$ orbitals. Fig. (\ref{Fig2} (b)) shows
that the molecular crystal field also lifts the degeneracy of the
one-dimensional bands closest to the Fermi energy, that are
associated with these two orbitals. We note that the splitting is
of about 0.5 eV. The figure also shows that these two bands are
exchange-splitted by an amount of 1.3 eV for the $d_{xz}$ and 0.4
eV for the $d_{yz}$. We take now the spin-up and down $d_{xz}$
bands and realize that they may be interpreted as the mean-field
ferromagnetic solution of a one-dimensional nearest-neighbors
Hubbard model at half-filling, with a hopping integral of 0.06 eV
(bandwidth of 0.25 eV), provided that there are no interband
excitations (e.g.: electrons from the filled majority-spin,
$d_{xz}$ band can not be excited to neither of the unfilled
majority or minority spin, $d_{yz}$ bands). The coulomb integral
$U$ can then be extracted by noting that the spin-splitting is
equal to $U n$, where $n=1$ is the filling fraction. Hence $U$ is
equal to 1.3 eV. But the ground state of a half-filled Hubbard
model is not a ferromagnetic metal, but an insulating
antiferromagnet. We have therefore simulated helical arrangements
of spins\cite{Gar04a,Gar04b}. We show in Fig. (\ref{Fig3} (a)) the
energy of these spin spirals as a function of the wave-vector $q$
of the helix. The figure shows that the ground state is indeed an
antiferromagnet. This fact further confirms our belief that these
chains are perfect playgrounds of Hubbard physics. A final word of
caution should however be added, since according to the former
arguments, the empty spin-up and -down $d_{yz}$ bands should not
be splitted at all.


\section{Chains of TMCp$_2$ molecules inside carbon nanotubes \label{SecIV}}

\subsection{Motivation and notation}

Isolated metallocene chains clearly lack mechanical stability. We
therefore propose that encapsulating them inside carbon nanotubes
should provide the wanted isolation and stabilization. Inclusion
of organic molecules inside carbon nanotubes entails however a
series of stringent requirements. Encapsulation of individual
molecules can surely be achieved if the chemical process is
exothermic. But assumed this condition is fulfilled, the energy
barriers for movement of the molecules along the axis of the
nanotube should be tuned to be not too high to completely prevent
this motion, nor too low to allow the molecules to displace
freely. Notice that in this last case molecules should leave the
tube under slight mechanical shake-ups. Chain formation from the
individual encapsulated molecules is helped by the own binding
energy of the chain. This additional source of stabilization
energy should link molecules together, and counteract the possible
negative effect of the detuning of the energy barriers.
Endothermic encapsulation of fullerenes has also been achieved
\cite{Lee02}. We therefore believe that individual metallocene
molecules may be encapsulated by endothermic processes.

\begin{figure}
\vspace{0.5cm}
\includegraphics[width=0.9\columnwidth]{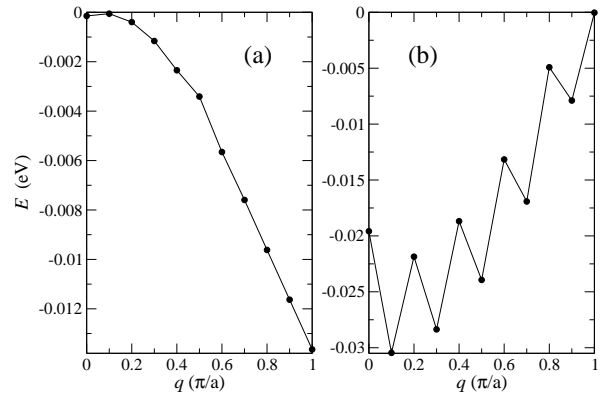}
\caption{\label{Fig3} Energy as a function of the spiral pitch
vector for (a) an isolated chain of cobaltocenes and (b) a chain
of cobaltocenes inside a (7,7) SWNT grown along the $C_5$ symmetry
axis. $q=0$ corresponds to the ferromagnetic configuration and
$q=1$ (in units of $\pi/a$) to the antiferromagnetic
configuration.}
\end{figure}

Encapsulation produces mechanical and electrical isolation of the
metallocene chain. It also dopes the nanotube and produces local
modifications of its band structure\cite{Lee02}. These local
modifications can be seen as electrostatic wells that trap
electrons. We expect these traps to be deepest for endothermic
encapsulation since in such a case the geometry would enhance the
interaction between molecules and nanotube walls. Notice also that
the length of the traps along the nanotube should be small for
metallic nanotubes since in this case the screening length is
pretty short. On the contrary, the length of the traps should be
much larger for semiconducting nanotubes.


We wish to stress the importance of commensurating the lattice
constant of the isolated chain (7.50 \AA) with that of the
nanotube, so that molecules accommodate well inside the tube and
the chain is neither compressed nor strained. If the metallocene
molecules are too close, electron clouds of two neighboring
molecules will overlap too much and the full device is rendered
endothermic. If they are too separated, then their wave functions
won't overlap appreciably and the molecules can be considered as
isolated. Since the unit cell of a armchair nanotube has a lattice
constant of 2.50 \AA, a metallocene may be well accomodated every
three nanotube unit cells. The lattice constant of zigzag naotubes
is 4.32 \AA, so placing a molecule every two nanotube unit cells
renders a stretched chain. Chiral tubes like the (12,4) or (15,5)
have a lattice constant of 15.59 \AA, and therefore one can safely
place two metallocenes inside each of them to form a somewaht
stretched chain. To avoid confusion, we use the following notation
to denote the basic unit cell of the simulations:
$\#$TMCp$_2@N(n,n')$, where $\#$ is the number of metallocenes
(TMCp$_2$) that are included in $N$ unit cells of a $(n,n')$ SWNT.
We have also placed the molecules in two possible orientations,
which we call parallel or perpendicular depending on whether the
$C_5$ axis of the nanotube is aligned parallel or perpendicular to
the nanotube axis.


\begin{figure}
\includegraphics[width=0.9\columnwidth]{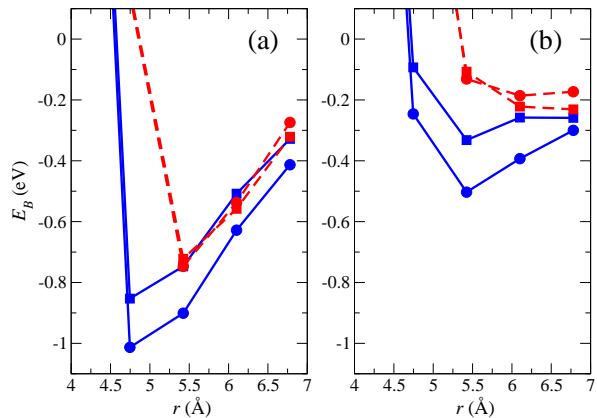}
\caption{\label{Fig4} Cohesive energy for the cases $N=3$
(circles) and $N=4$ (squares) in the parallel (continuous line)
and perpendicular (broken line) configurations, calculated without
(a) and with (b) BSSE correction.}
\end{figure}

\subsection{Metallic nanotubes}

We have found that the BSSE artificially enhances the stability of
the encapsulated chains, so that BSSE corrections are essential to
get right answers. This is explicitly shown in Fig. (\ref{Fig4})
for CoCp$_2$ chains, encapsulated in armchair tubes of different
diameter. Notice that the most stable configuration corresponds to
a (8,8), or (7,7) nanotubes (radii of 5.4 or 4.7 \AA,
respectively) depending on whether BSSE is corrected or not.
Hence, we shall only discuss our results with BSSE correction,
Fig. (\ref{Fig4} (b)). We note that encapsulation is exothermic
for armchair $(n,n)$ tubes, with $n$ equal or larger than 7 (radii
equal or larger than 4,7 \AA). The figure predicts that the
highest rate of encapsulation should be achieved for (8,8) tubes,
with smaller rates for (7,7), (9,9), (10,10) and so on. But the
TEM images of Li {\em et al.} \cite{Li05} only show encapsulation
for radii similar to that of the (7,7) or slightly larger, and
therefore predict a very selective encapsulation. We believe this
is due to the small energy barriers that a metallocene would find
when it moves along the axis of thicker tubes. Since experiments
are performed at room temperature and samples are transported
between different places, small barriers wouldn't prevent the
metallocene to escape when mechanical and thermal perturbations
increase its mobility.

Before switching to the analysis of the electronic structure we
shall further analyze the results presented in Fig. (\ref{Fig4}
(b)). The figure demonstrates that molecules fit inside (7,7)
tubes only if their axes are aligned. Misaligned molecules can be
placed inside thicker tubes, but with a pretty large energy cost.
We shall therefore only analyze aligned metallocene molecules in
what follows. According to the same figure metallocenes can also
lower significantly their energy by sticking close enough, which
means that the most stable configuration corresponds to a
molecular chain inside the nanotube. This is also confirmed
experimentally by the TEM images of Li {\em et al.} \cite{Li05},
where the cobaltocenes show a clear tendency to appear together.

\begin{figure}
\includegraphics[width=0.8\columnwidth]{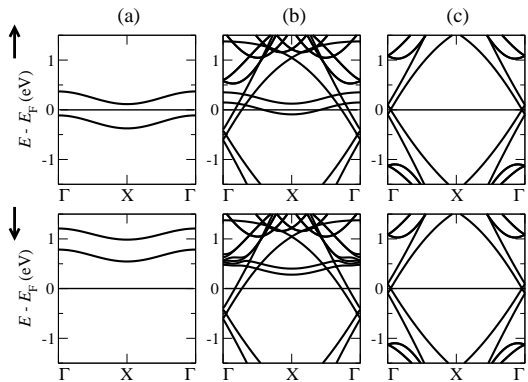}
\caption{\label{Fig5} Band structure of (a) an isolated
cobaltocene chain (lattice constant = 7.5 \AA), (b) a cobaltocene
inside 3 unit cells of a (7,7) nanotube (the same separation
between molecules as before) and (c) a (7,7) nanotube, for spin-up
and spin-down electrons.}
\end{figure}

We turn now to discuss the electronic properties of the preferred
CoCp$_2$@3(7,7) devices. The first remarkable property of these
systems is a charge transfer of 0.6 $e^-$ from the cobaltocene to
the nanotube, which is due to the high electron affinity of the
latter \cite{Lu04}. Such a charge transfer, which is responsible
for the reduction of the magnetic moment of the molecule from 1 to
0.4 $\mu_B$, can also be inferred from the shift in the position
of the spin-up $d_{xz}$ band when the chain is inserted in the
tube. This shift can be appreciated by comparing the band
structures of isolated and encapsulated chains, that we depict in
Figs. (\ref{Fig5} (a), (b)). This transfer makes the valence bands
of the nanotube move down, as can be seen in Figs. (\ref{Fig5}
(b), (c)). On the other hand the crystal field splitting of the Co
$d$ bands decreases from about 0.5 to 0.2 eV. This effect is
mainly due to the change in the atomic positions of the
encapsulated cobaltocene. Indeed, by plotting the bands of an
isolated chain calculated using the coordinates of the cobaltocene
in the CoCp$_2$@3(7,7) structure we find that the change in the
crystalline field is very similar. The screening and charge
transfer also produces a large reduction of the Stoner splitting
from 1.3 to 0.5 eV. Taking the mean field estimate for this
splitting, $\Delta E \approx n\,U = 0.4\,U$, gives a Hubbard $U$
of 1.3 eV again. We therefore infer that screening effects
produced by the nanotube do not reduce the Coulomb interaction in
these chains.

\begin{figure}
\includegraphics[width=0.8\columnwidth]{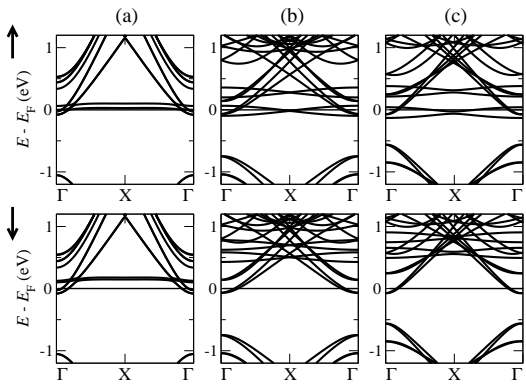}
\caption{\label{Fig12} Band structure of (a) CoCp$_2$@2(11,0) (b)
2CoCp$_2$@(12,4) and (c) 2CoCp$_2$@(15,5) for spin up and spin
down electrons.}
\end{figure}

The filling of the metallocene chains modifies the mean-field
ground state, that was insulating and antiferromagnet in vacuum,
to an helical state with a small pitch vector very close to the
ferromagnetic solution, as can be appreciated in Fig. (\ref{Fig3}
(b)). The small oscillations in the energy are due to the
application of the spiral rotation to the nanotube states.


\subsection{Semiconducting nanotubes}

We extend now the same discussion to the case of semiconducting
nanotubes. The evolution of the binding energy and charge transfer
to the nanotube wall as a function of the nanotube radius is very
similar to the armchair case. Zigzag nanotubes start being
exothermic for a radius of $\sim 4.7$, which corresponds to the
(12,0) tube. The chiral nanotubes (12,4) and (15,5), with radii of
5.7 and 7.2 \AA\/ respectively, have also been found to be
exothermic. Furthermore, we have found that the charge transfer
decreases as the radius of the nanotube increases, as in armchair
tubes.

We show in Fig. (\ref{Fig12}) the band structures of
CoCp$_2$@2(11,0), 2CoCp$_2$@(12,4) and 2CoCp$_2$@(15,5) as
representative of stretched and reasonably commensurated chains.
The most noticeable feature in the first case, which corresponds
to an endothermic encapsulation, is the severe reduction of the
crystalline field between the $d_{yz}$ and $d_{xz}$ bands as a
consequence of the change in the atomic positions produced by the
strong interaction with the nanotube wall. Note also that these
bands are almost non-dispersive due to the large separation
between metallocenes along the nanotube axis, which are placed
about 8.64 \AA apart. The chiral cases are very similar, despite
the large difference between their radii, although, as expected,
the charge transfer is larger in the (12,4). In both cases the
cobaltocene bands are clearly dispersive and show a crystalline
field splitting which in the (15,5) is slightly larger due to the
smaller interaction with the nanotube wall. Note that bands
associated wit the cobaltocene are folded due to the presence of
two molecules in the unit cell.


The Hubbard $U$ can not be cleanly estimated in the endothermic
(11,0) encapsulation due the strong modification of the
cobaltocene bands and the hybridization with the nanobute states.
However, in the chiral nanotubes by taking into account the
exchange splittings, which are 0.72 and 0.87 eV in the (12,4) and
(15,5), respectively, and the occupations of each spin up $d_{xz}$
level, which are 0.57 and 0.67 electrons, respectively, $U$ turns
out to be again 1.3 eV, in agreement with the armchair nanotube.
This demonstrates the robustness of the calculation.

\section{Mapping to strongly correlated models \label{SecV}}

\subsection{Physics at the chain}

The discussion following Figs. (\ref{Fig2} (b)) and (\ref{Fig5}
(a), (b)) highlighted the similarities of the band structure,
ground state and magnetism of isolated or encapsulated metallocene
chains to strongly correlated Hubbard chains at diverse doping
levels. We wish to take a closer look to this delicate issue now.

Indeed, an isolated  metallocene chain may be viewed as a series
of magnetic impurities stacked on top of each other, whose
wave-functions have little overlap. The result of the process of
encapsulation may therefore be a sort of Kondo chain, where the
metallocene chain acts a an impurity lattice that hybridizes with
the electronic (Luttinger) liquid residing in the host nanotube.

The parameters that tune the model are therefore the
hybridizations among metallocene molecules $V_{MM}$, or between
metallocenes and the nanotube $V_{MN}$. We note that $V_{MM}$ is
rather small for many of the isolated or encapsulated metallocene
chains since these molecules are fairly stable and therefore do
not like to share electrons. Exceptions are CoCp$_2$ and NiCp$_2$,
whose bandwidth are sizeable. We shall therefore confine  the
discussion below again to cobaltocene chains. In order to estimate
the size of the nanotube-molecule hybridization integral $V_{MN}$,
we first look into the band structure of the composite device in
Fig. (\ref{Fig5} (b)). The figure shows that the most important
effect of encapsulation is a transfer of charge that shifts the
metallocene bands upwards and the nanotube bands downwards, and,
associated with it and with a change of atomic positions, a
reduction of crystal field and exchange splittings. Nanotube and
metallocene bands close to the Fermi level cross each other with
no apparent hybridization. This fact indicates that $V_{MN}$ is
essentially zero. Additional confirmation comes from the fact that
nanotube bands in the encapsulated device are not spin-splitted
and therefore the nanotube does not magnetize.

\begin{figure}
\includegraphics[width=0.9\columnwidth]{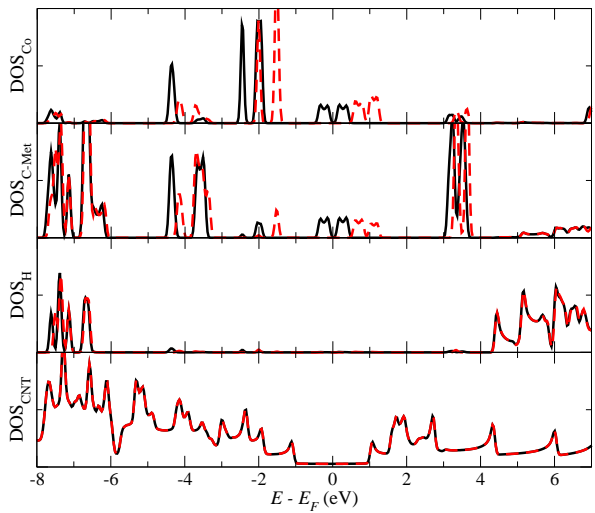}
\caption{\label{Fig6} From top to bottom, projected densities of
states on the cobaltocene Co, C and H atoms for a Cobaltocene
chain at its equilibrium distance, and a clean (7,7) SWNT. Solid
and dashed lines indicate spin up and down components,
respectively.}
\end{figure}

\begin{figure}
\includegraphics[width=0.9\columnwidth]{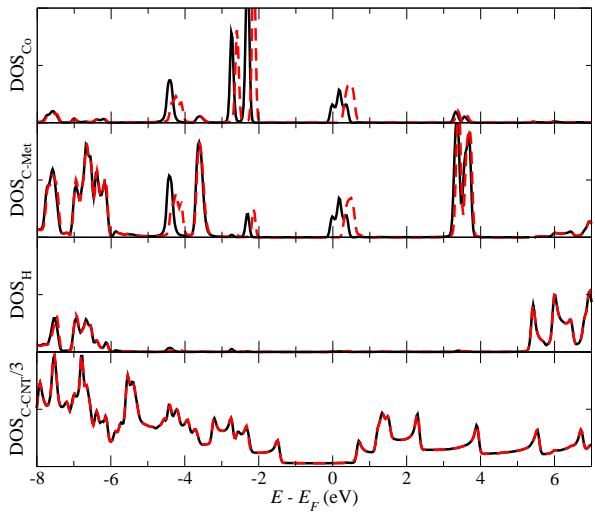}
\caption{\label{Fig7} From top to bottom, projected densities of
states on the cobaltocene Co, C and H atoms and the naotube walls
in a CoCp$_2$@3(7,7) device. Solid and dashed lines indicate spin
up and down components, respectively.}
\end{figure}

\begin{figure}
\includegraphics[width=0.9\columnwidth]{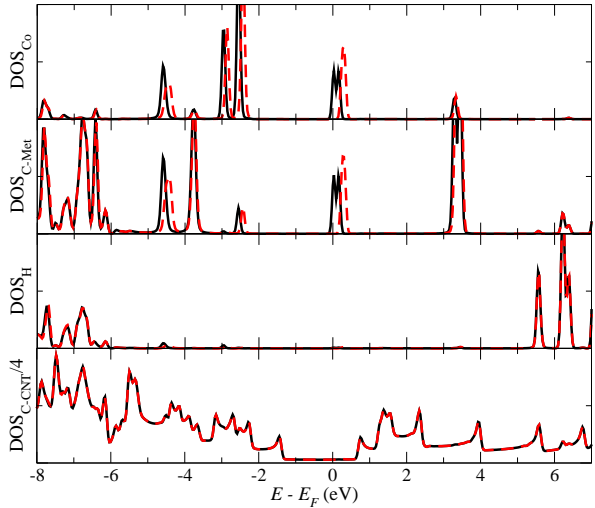}
\caption{\label{Fig8} From top to bottom, projected densities of
states on the cobaltocene Co, C and H atoms and the naotube walls
in a CoCp$_2$@4(7,7) device. Solid and dashed lines indicate spin
up and down components, respectively. }
\end{figure}

We cross-check this estimate of the hybridization by taking a look
at the projected density of states (PDOS) of the different
molecule and nanotube atoms, which is plotted in Figs.
(\ref{Fig6}), (\ref{Fig7}) and (\ref{Fig8}) for the cases of an
isolated CoCp$_2$ and a (7,7) nanotube, a CoCp$_2$@3(7,7) chain
and a CoCp$_2$@4(7,7) chain, respectively. All of these devices
have been simulated for the same lattice constant, which is that
of the nanotube. Comparison of Figs. (\ref{Fig6}) and
(\ref{Fig7})/(\ref{Fig8}) confirms our view that the main effect
of encapsulation is a transfer of charge and a structural
modification that brings about a reduction of crystal field and
exchange splittings. Furthermore, the last panel in Figs.
(\ref{Fig7}) and (\ref{Fig8}) shows no noticeable splitting of the
spin-resolved DOS projected on the carbon atoms of the nanotube.

We further notice that the shape of the DOS of cobalt atoms does
not change too much. The only noticeable changes appear in the
energy range [-8,-6] eV below the Fermi energy in the DOS
projected on the hydrogen and carbon atoms. What remains to be
seen is whether these modification are due to structural
deformation of the molecules and nanotube walls after
encapsulation, or to hybridization between them. We note that the
sharp resonances associated to hydrogen atoms of the cobaltocene
chain are clearly broadened after encapsulation. We therefore
believe that Hydrogen atoms, and partly carbon atoms, do
hybridize. This hybridization does have no implications in the
electronic properties of the devices, since they appear well away
the Fermi level. They provide a glue between molecules and tube,
that hinders the axial motion of the molecule.

\subsection{Physics at the walls}

We now look at the change in the electrostatic potential at the
nanotube walls created by the combined effect of charge transfer,
structural modification, and hybridization with hydrogen atoms. We
expect that the potential will show well-defined wells around the
positions where the metallocene molecules are placed, that will be
better or worse screened depending on the metallice or
semiconducting character of the walls. These wells, that should
trap electrons \cite{Lee02,Cho03}, will form a periodic chain with
the same lattice constant as the underlying metallocene chain. A
good estimate of the shape, length and depth of these wells, can
be provided by calculating the electrostatic potentials at the
walls of the nanotube of a CoCp$_2$@$N(n,n')$ device with and
without metallocene molecules, and substracting one from the
other, $\Delta V(z)=V_{H,with}(z)-V_{H,without}(z)$. We note that
this potential should have axial symmetry and hence only depend on
the $z$-coordinate.

\begin{figure}
\includegraphics[width=0.9\columnwidth]{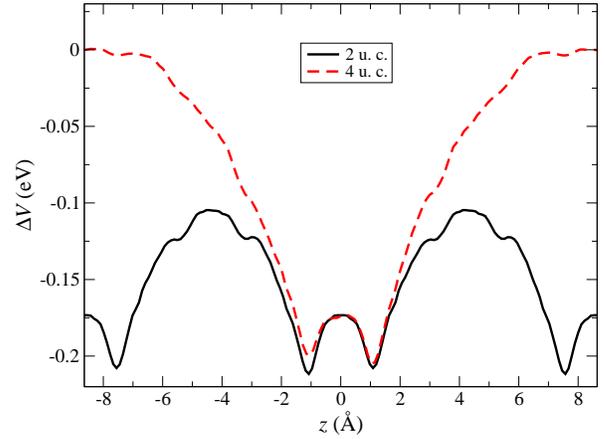}
\caption{\label{Fig9} $\Delta V$ as a function of the
$z$-coordinate for CoCp$_2$@2(11,0) (continuous line) and
CoCp$_2$@4(11,0) (dashed line) devices.}
\end{figure}

We have checked that $\Delta V$ is rather small for metallic
devices since, as expected, the valence electrons at the walls
screen the charge perturbation efficiently. We have subsequently
looked into the semiconducting CoCp$_2$@$N$(11,0) device (radius
of 4.4 \AA), whose band structure is plotted in Fig. (\ref{Fig12}
(a)). In this case the encapsulation is endothermic and therefore
the modification of the potential is expected to be further
enhanced as a consequence of the strong interaction between the
metallocene and the nanotube wall. Indeed we have found that the
potential barrier is rather large as can be seen in Fig.
(\ref{Fig9}), where we plot $\Delta V$ as a function of the axial
coordinate for a device where $N=2$ and for another where $N=4$.
We note that we have placed one molecule at the origin of the
$z$-coordinate. We inmediately note the periodicity in $\Delta V$
for both $N=2$ (8.6 \AA) and $N=4$ (17.2 \AA). The potential has
two symmetric minima, placed at the positions of the close-by
$C_5$ rings and a maximum in between them, that corresponds to the
position of the cobalt atom, which is farther apart. The absolute
maximum in $\Delta V(z)$ appears at the middle point between two
molecules. The height of the well depends on both the lattice
constant of the chain and the screening length. The figure shows
that $\Delta V(z)$ has enough room to reach zero when $N=4$,
allowing us to estimate a value of about 6-7 \AA\/ for the
screening length. For $N=2$, on the contrary, the distance between
wells (8.6 \AA) is not long enough to allow the complete healing
of the potential. Accordingly, the depth of the well decreases
from 0.2 to 01 eV.


\section{Transport Properties of Isolated Metallocenes Inside Carbon
Nanotubes \label{SecVI}}

\begin{figure}
\includegraphics[width=0.9\columnwidth]{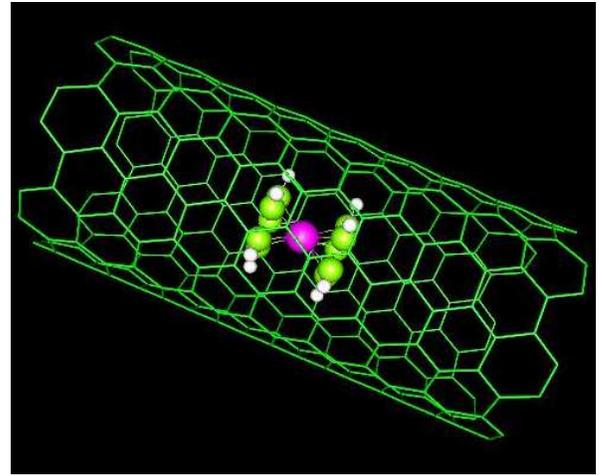}
\caption{\label{Fig10} Unit cell used to calculate the transport
properties of a clean nanotube with a metallocene impurity. The
atoms of the metallocene have been highlighted for clarity.}
\end{figure}

The case where cobaltocenes form a chain inside the nanotube
\cite{Gar06} can lead to magnetorresistive ratios of $20\%$ due to
the suppression of the spin channel that goes through the
molecular chain. A different situation however is that produced
when a cobaltocene sits in between two infinite nanotube
electrodes, as shown in Fig. (\ref{Fig10}).

\begin{figure}
\includegraphics[width=0.9\columnwidth]{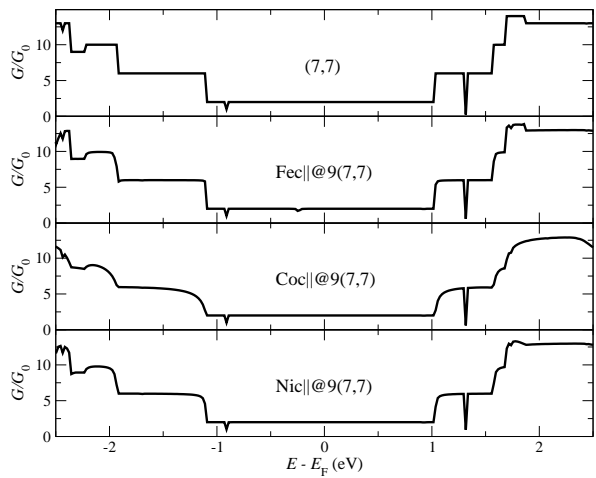}
\caption{\label{Fig11} Transmission coefficients of isolated
ferrocenes, cobaltocenes and nickelocenes in the parallel
configuration between semi-infinite (7,7) SWNT leads. For
comparison it is also shown the transmission coefficients of a
clean (7,7) SWNT.}
\end{figure}

The conductance of such a system is plotted in Fig. (\ref{Fig11})
for three different metallocenes, namely ferrocene, cobaltocene
and nickelocene in the parallel configuration. These are
representative cases since the other metallocenes, with the
possible exception of the ruthenocene and the osmocene, have
smaller dimensions and are expected to produce less interaction
with the nanotube wall. The transmission of the clean nanotube is
also plotted  for comparison. As can be seen the net effect
produced on the transmission coefficients by isolated metallocenes
is not very large and the transmission at the Fermi level is not
affected. Notice that in the case of the magnetic molecules the
transmission is exactly the same for spin-up and spin-down
electron due to the fact that the carbon nanotube does not acquire
any spin polarization from the charge that is transferred from the
metallocene. The only differences with a clean nanotube appear
above and below the fermi level and have the effect of rounding
the transmission coefficients where the number of channels
increases. This effect is mainly due to the structural deformation
induced by the metallocene rather than to any electronic
influence, although it could also be due to the charge transfer.
The change seems to be very small for the ferrocene impurity
whereas the cobaltocene produces the largest modifications.

We therefore predict that there should be no change in the
transport measurements of isolated metallocenes inside nanotubes
with radii larger 4.5 \AA, unless the applied voltage is large
enough ($>2$ V) to reach the regions with large number of
channels. In such a case the presence of scattering would decrease
the measured current and conductance and this effect would be
clearly seen in the second derivative of the current with respect
to the voltage. It is also worth mentioning that such and effect
would be more pronounced in nanotubes with small diameter but the
encapsulation of metallocenes would be difficult to realize due to
the endothermic character, which would probably produce defects
that would have a larger influence on the transport properties.


\section{Conclusions}

We have shown that density functional theory in combination with
pseudopotentials and linear combinations of atomic orbitals
provides a very accurate description of the electronic and
structural properties of metallocenes, whose results are in
excellent agreement with experiments. The electronic configuration
of metallocenes, which is mainly determined by the interplay
between the crystalline field splitting and Hund's rules on the
$d$ bands, shows a kind of mirror symmetry with respect to the
column of iron due to the number of unpaired electrons. The
filling of the $d$ bands and the shape and symmetry of the
orbitals involved can also explain the distribution of the
structural parameters as a function of the repulsion between the
orbitals of the cyclopentadienyl rings and those of the transition
metal atom. As a consequence the metallocenes become bigger as the
number of electrons of the metallic atom increases.

The electronic configuration of metallocene chains does not differ
too much from that of the isolated metallocenes, the main
difference being the finite dispersion of the states around the
Fermi level, which make them conducting. Such states, which in the
cobaltocene are half-filled are responsible for the magnetic
configuration of the chains, which we demonstrated to be
antiferromagnetic by calculating the energy of a series of helical
configurations with the help of the generalized Bloch's theorem.
The spin polarization of those states makes these systems perfect
candidates for spintronics applications. However, it is necessary
to look for possible methods to stabilize them without destroying
such interesting properties.

The most obvious way of producing stable metallocene chains is by
placing them inside a carbon nanotube, which would act as an
external coating that would protect them mechanically and
electrically. Such a system was found to be exothermic for radii
larger than 4.5 \AA, in agreement with recent experiments. The
encapsulated metallocenes produce a transfer to the nanotube wall
and the carbon nanotube becomes $n$-doped. This effect is clearly
seen when the band structure or the density of states of the
composite system is analyzed. Such charge transfer changes also
the most stable magnetic configuration which moves close to the
ferromagnetic point with a small spiral pitch vector, although the
difference in energy is small enough to point all spins up with
the application of a magnetic field. By closely analyzing the
projected density of states it is also possible to study the
interaction of the molecular magnetic moment with the conduction
electrons of the carbon nanotube.

Our mean-field calculations show that CoCp$_2$ chains behave as
strongly correlated electron liquids, that can be perfectly well
described by a half-filled Hubbard model. We further note that the
hopping $t$ and Coulomb $U$ parameters can be univocally extracted
from our simulations. We also show that encapsulation only
produces a charge transfer that carries the Hubbard liquid away
from half-filling.

Looking into the effects on the nanotube walls, we have shown that
encapsulation inside semiconducting nanotube produces deep wells
in the electrostatic potential that should act as quantum dots
that trap electrons.

Finally, we also studied the transport characteristics of
metallocenes between semi-infinite clean nanotubes. We found that
the transmission coefficients do not change too much with respect
to the transmission coefficients of a clean nanotube, the only
difference being the rounding of the steps where the number of
channels increases. This would give rise to changes in the second
derivative of the current with respect to the voltage for large
bias voltages.

\begin{acknowledgments}
We would like to thank John H. Jefferson for useful and
stimulating discussions. We acknowledge financial support from the
UE network MRTN\--CT\--2003\--504574 RTNNANO, the Spanish project
MEC BFM2003-03156 and the UK EPSRC. VMGS thanks the European Union
for a Marie Curie grant.
\end{acknowledgments}

\end{document}